# Control of spiral waves and turbulent states in a cardiac model by travelling-wave perturbations

Peng-Ye Wang[*], Ping Xie, and Hua-Wei Yin

*Laboratory of Soft Matter Physics, Institute of Physics,  
Chinese Academy of Sciences, Beijing 100080, China*

**Abstract**

We propose a travelling-wave perturbation method to control the spatiotemporal dynamics in a cardiac model. It is numerically demonstrated that the method can successfully suppress the wave instability (alternans in action potential duration) in the one-dimensional case and convert spiral waves and turbulent states to the normal travelling wave states in the two-dimensional case. An experimental scheme is suggested which may provide a new design for a cardiac defibrillator.

## 1. Introduction

Since the first report of Beck et al. [1], delivering an electric shock to reset the heart has become a common effective clinical method to save the lives of patients suffering from ventricular fibrillation, which is the most common arrhythmia associated with sudden cardiac death. Such an electric shock depolarizes all heart cells simultaneously and interrupts the chaotic twitching of heart muscle. However, this defibrillation process may cause damage to the cardiac tissue because of the high discharge energy of the defibrillator [2]. The damage can make ventricular fibrillation more likely to reemerge. Furthermore, strong defibrillation shocks are painful if applied to a patient who is still conscious. Therefore, finding a new way of defibrillation avoiding strong massive electric shock is highly desirable and of great importance.

Over the last decade, it has become recognized that one possible mechanism responsible for the sudden transition from ventricular tachycardia to fibrillation is the spontaneous breakup of the spiral wave of action potential into a spatiotemporal turbulent state [3-8]. Great efforts have been focused on the temporal control of instability in an *in vitro* slice of rabbit interventricular septum tissue [9] and in the artrioventricular nodal system [10-12]. In a model of mammalian ventricular tissue, Biktashev and Holden [13] used small amplitude, spatially uniform repetitive stimulation to push the spiral wave to the side in order to eliminate it. Rappel, Fenton and Karma [14] suggested a time-delayed feedback control scheme to control the spatiotemporal chaotic state in the one-dimensional (1D) Beeler-Reuter model [15] and to prevent spiral wave breakup in a two-dimensional (2D) excitable cardiac model [5]. A limiting factor of the method in Ref. [14] is that the control must be achieved before the spiral wave breakup occurs. In addition, spatial perturbation method [16] has been shown to be an effective way in controlling spatiotemporal dynamics in non-excitable systems. In this work, we present a travelling-wave perturbation method to control wave instabilities in an excitable cardiac model. With our method, both the turbulent wave, generated from the breakup of the spiral wave, and the spiral itself, which is considered to correspond to ventricular tachycardia that is also an abnormal state of the heart, can be eliminated effectively. Furthermore, our technique can be easily realized in practice because no feedback is necessary. A detailed design of the control scheme is given. With this scheme we present a possibility of applying a travelling-wave electric field perturbation to the ventricular myocardium. We anticipate that the amplitude of the perturbation is much smaller than the magnitude of the intrinsic action potential pulse of the cardiac tissue. Therefore, the "defibrillator" would not cause any damage to the heart.

---

[*] Author to whom correspondence should be addressed. E-mail: pywang@aphy.iphy.ac.cn

**2. Model and results**

To demonstrate our technique, we consider the two-variable excitable cardiac model [5]:

$$\varepsilon \frac{\partial E}{\partial t} = \nabla^2 E - E + \left[ A - \left( \frac{n}{n_B} \right)^M \right] [1 - \tanh(E - 3)] \frac{E^2}{2} + E_P(\mathbf{r}, t), \quad (1a)$$

$$\frac{\partial n}{\partial t} = \theta(E - 1) - n, \quad (1b)$$

where E is the transmembrane potential, n is the slow current gate variable, $\theta(x)$ is the standard Heaviside step function, $\varepsilon$ is the relaxation time characterizing the abruptness of excitation, and M is a parameter controlling the wave-front insensitivity. Throughout the calculation we fix A=1.5415, and $n_B$=0.507 as in Ref.[5]. We choose this model because of its advantages: it circumvents the complexity of cell electrophysiology while still making contact quantitatively with cardiac tissue properties, and it is easier to perform numerical simulations than with the multi-variable electrophysiological models such as the Noble model [17].

The term $E_P(\mathbf{r},t)$ on the right hand side of Eq.(1a) is an externally applied voltage, where **r** is a vector denoting spatial coordinates. The purpose of using $E_P(\mathbf{r},t)$ is to force the cardiac tissue to the normal state. Considering that the normal excitation state of the cardiac tissue is a travelling wave and reminding our previous works on controlling spatial patterns [16] where the perturbation mimics the target pattern, we choose the control voltage $E_P(\mathbf{r},t)$ as a travelling wave function. It is anticipated that this form of perturbation can effectively eliminate the irregular electrical wave propagation (spiral waves and turbulent states) in the cardiac tissue and convert the heart to the normal state. The amplitude of $E_P(\mathbf{r},t)$ should be as small as possible in order that the perturbation has the least damage to the cardiac tissue. Except that the spatial period and temporal frequency of the perturbation function should be consistent with the wave vector and the velocity of the normal cardiac travelling wave, respectively, the exact form of $E_P(\mathbf{r},t)$ is insignificant, because after the system is excited the shape of the excitation pulse is only relevant to that obtained from Eq.(1) without the external perturbation, provided that the amplitude of $E_P(\mathbf{r},t)$ is much smaller than that of the action potential pulse. The simplest form of $E_P(\mathbf{r},t)$ satisfying the above requirement is a sinusoidal function, i.e.,

$$E_P(\mathbf{r},t) = \alpha \cos(\mathbf{k} \cdot \mathbf{r} - \omega t) \tag{2}$$

where $\alpha$ is the amplitude of the perturbation, $\mathbf{k}$ is the spatial wave vector with its magnitude and direction determining the spatial separation and the propagating direction of the target wave pulse, respectively, and $\omega$ is the frequency determining the repetition rate (or moving speed) of each pulse. As mentioned above, $\alpha$ should be much smaller than the amplitude of the action potential pulse but must be larger than a certain threshold $\alpha_{th}$ in order to realize effective perturbation.

Starting from the zero-dimensional (0D) case, we drop the spatial Laplacian term in Eq.(1). The nullcline structure (without perturbation) is shown in Fig.1(a). The excitable fixed point A is $(E_0, n_0) = (0, 0)$. For the case of $E < 1$ (both points A and B in Fig.1 satisfy this condition), equation (1) can be approximately written as follows:

$$\varepsilon \frac{\partial E}{\partial t} = -\frac{\partial f(E)}{\partial E} + \alpha \cos(\omega t), \tag{3}$$

where $f(E) = \frac{1}{2}E^2 - \frac{1}{3}AE^3$ is a cubic potential function as shown in Fig.1(b). With this approximation we can obtain the excitation threshold for E analytically, i.e., $E_B = \frac{1}{A} \approx 0.6487$, while from the calculation of E-nullcline we get $E_B \approx 0.6547$, which shows good agreement with each other. From this simple analysis we can see that the perturbation acts as an external driving force to the system moving along the potential curve as shown in Fig.1(b). If the amplitude $\alpha$ of the perturbation is very small, the system will oscillate around the stable fixed point A with small oscillation amplitude. When we increase $\alpha$ to a certain threshold $\alpha_{th}$ the perturbation can push the oscillation to reach the unstable fixed point B. Over this point the cardiac tissue is excited by the perturbation. Naturally, in order to control the global behavior (not local behavior around a fixed point), we need the perturbation amplitude $\alpha > \alpha_{th}$. The inset of Fig.1(b) shows the value of $\alpha_{th}$ versus $\omega$, calculated with Eq.(1) for $\varepsilon = 0.05$ and $M = 4$. The relation between $\alpha_{th}$ and $\omega$ is not dependent on the value of M because n is always zero before the excitation. We can see that $\alpha_{th}$ increases with $\omega$ almost linearly. From the above analysis we conclude that if the perturbation frequency $\omega$ is not too large (say less than 6), small perturbation amplitude (say 0.25), which is much smaller than the magnitude of the action potential pulse, is enough to control the global behavior of the cardiac tissue.

Let us first extend the perturbation scheme to spatiotemporal control of 1D waves. It is known [5] that alternans in action potential duration (APD) occur due to the oscillatory pulse instability which is an abnormal state of the cardiac tissue. The purpose of our control scheme is to suppress the wave instability and, therefore, to regulate the APD to a constant. With the travelling-wave perturbation of Eq.(2), we use a split step spectral method to numerically integrate Eq.(1) in the 1D line segment with a periodic boundary condition. The line segment of length $L = 16\pi$ is uniformly divided into 512 parts for performing Fourier transformation in order to calculate the Laplacian term for each integration time step (dt = 0.01). A typical result demonstrating the control in 1D is shown in Fig.2. We find that the spatiotemporal chaotic oscillatory pulses are stabilized, resulting in a regular travelling wave with the same velocity direction as the control signal. The transient process is only in a few oscillation numbers, indicating a very fast control.

Figure 3(a) shows the controllable region in the plane of the control parameters k and ω. Due to the fact that the spatial period and temporal frequency of the target state is the same as those of the control signal, as mentioned above, the wave vector and the velocity of the desired final travelling wave can be freely manipulated inside the controllable region. The controllable region is larger when the amplitude of the perturbation is increased. Figure 3(b) shows the value of the threshold controlling amplitude $\alpha_{th}$ versus k. Control is successful only when $\alpha \geq \alpha_{th}$. We find that there exists an optimum value of k ( ~ 0.9 ) where $\alpha_{th}$ is the smallest ( ~ 0.22 ).

The preceding 1D results provide the basis for controlling spiral waves and turbulent states in 2D. In the 2D case, numerical simulations are performed using the same split step spectral method as in the 1D case on a $512 \times 512$ grid with a box size of $32\pi$. The integration time step is taken to be dt = 0.01. Figure 4(a) shows the spiral wave in a region of $27 \times 27$. With the sinusoidal travelling wave perturbation of Eq.(2), moving upwards in the vertical direction the spiral wave is successfully eliminated and a stable travelling wave with the same wave vector and moving velocity as the perturbation signal is obtained, as shown in Fig.4(b). The oscillation structure is converted to the normal cardiac action potential pulses free from reentrant excitation. The turbulent state, shown in Fig.4(c), from the spontaneous breakup of the spiral wave is also effectively controlled to the normal state, Fig.4(d), with the sinusoidal travelling wave perturbation. We notice that a weak perturbation (compared with the pulse amplitude 3.8) is enough to realize the control. As in the 1D case, the control can be realized in a large range of the control parameters, e.g., when we take a k value half of that in Fig.4 and keep the other parameters the same, both the control of the spiral wave and the turbulent state in Fig.4 are still successful. We verified that our method can also effectively control a turbulent states that developed from the spiral breakup induced by temporary spatial inhomogeneities or obstacles rather than from the spontaneous breakup.

We emphasize that, after the spirals and turbulent states are eliminated and the travelling waves are obtained, when we remove the control signal the travelling waves persist stably and the spirals and turbulent states never show up again unless severe defects are caused. We define the threshold $\alpha_{th}$ for the amplitude of the perturbation in such a way that when $\alpha > \alpha_{th}$, the control is always successful, otherwise the spiral waves and the turbulent states cannot be completely eliminated. The threshold amplitude of the perturbation to control the spiral wave is around $\alpha_{th} = 0.25$. The threshold amplitude of the perturbation to control the turbulent states is around $\alpha_{th} = 0.28$.

The results above are for the sinusoidal travelling-wave perturbation. Other forms of the travelling wave may also be used as perturbation functions. For example, we can use square and triangular travelling waves, as shown in Figs.5(b) and (c), respectively. For comparison, the sinusoidal travelling wave is also plotted in the figure [Fig.5(a)].. Numerical simulations show that both the square and the triangular travelling waves also work very well. They can play the same role as the sinusoidal travelling wave does. Only the threshold amplitude is slightly different, e.g., for controlling the spiral wave as shown in Fig.4(a), we get $\alpha_{th} = 0.21$ for a square wave and $\alpha_{th} = 0.29$ for a triangular wave. For controlling the turbulent state as shown in Fig.4(f), we get $\alpha_{th} = 0.22$ for a square wave and $\alpha_{th} = 0.34$ for a triangular wave. These results are understandable because the square wave has a larger effective power than the triangular wave and the sinusoidal wave is between them if they have the same amplitude.

## 3. Experimental scheme

Based on the preceding analysis, we here present an experimental scheme which may provide a new design for a defibrillator. Parallel electrodes are uniformly placed in space as shown in Fig.5(d), where "−", "o", and "+" denote negative, zero, and positive voltages, respectively. These electrodes deliver a spatial voltage distribution which can be approximately plotted as a triangular wave, denoted by the solid dots in Fig.5(e). After a time interval of $\Delta t$, we change the voltage of each electrode to a new value as shown in Fig.5(f). Now the spatial voltage distribution becomes the triangular wave denoted by circles in Fig.5(e) which is moved $\frac{1}{4}$ spatial period rightward, respective to the previous one. If we repeat this procedure after the same time interval of $\Delta t$, a discontinuously moving travelling wave, having a spatial period of 4 times the electrode separation and a temporal frequency $\omega = \frac{\pi}{2\Delta t}$ is generated. Further detailed numerical simulations show that the perturbation of this discontinuous travelling wave has exactly the same effect as that of the continuous triangular travelling wave in the control. By changing the electrode separation and the time interval $\Delta t$, the defibrillator can be adjusted to fit the beating rate of the normal heart.

## 4. Conclusion

In conclusion, we have presented a travelling-wave perturbation method to control the spiral waves and turbulent states in an excitable cardiac model. Numerical simulations show that the perturbation can effectively eliminate the spiral waves corresponding to ventricular tachycardia and turbulent states corresponding to fibrillation in the heart. The weak perturbation provides a clue for the possibility to avoid strong massive electric shock in defibrillation. An experimental scheme is suggested which may provide a new design for a cardiac defibrillator.

**ACKNOWLEDGMENTS**
This research was supported by the National Natural Science foundation of China (Grant number 60025516).

Figures:

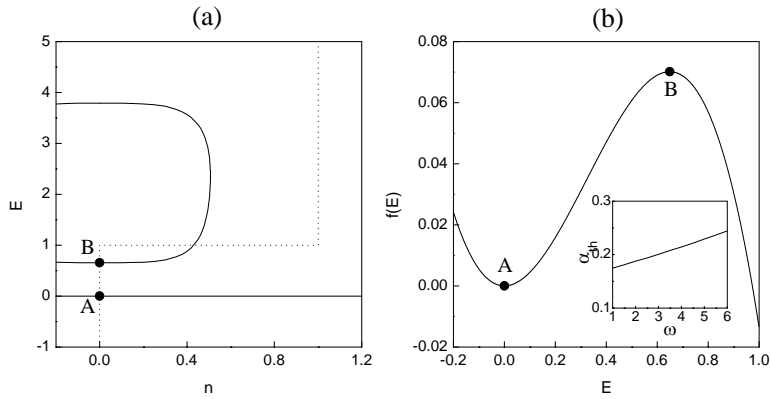

Fig.1. (a) E-nullcline (solid line) and n-nullcline (dashed line) for $M=4$. (b) Potential function $f(E) = \frac{1}{2}E^2 - \frac{1}{3}AE^3$. The inset shows the threshold of $\alpha_{th}$ versus $\omega$ obtained from Eq.(1) for $\varepsilon = 0.05$ and $M = 4$.

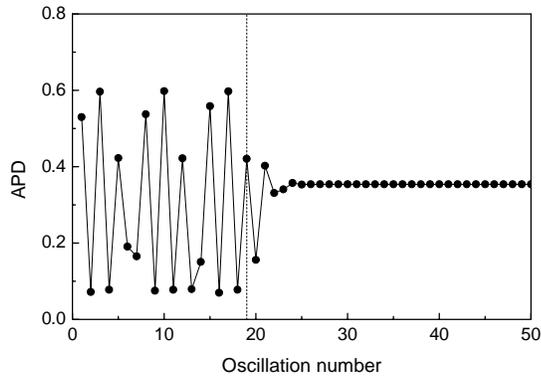

Fig.2. Action potential duration (APD) as a function of oscillation number in the 1D case. The vertical dashed line indicates the oscillation number when the control is turned on. $\varepsilon = 0.1$, $M = 10$, $k = 1$, $\omega = 4.6$, and $\alpha = 0.32$.

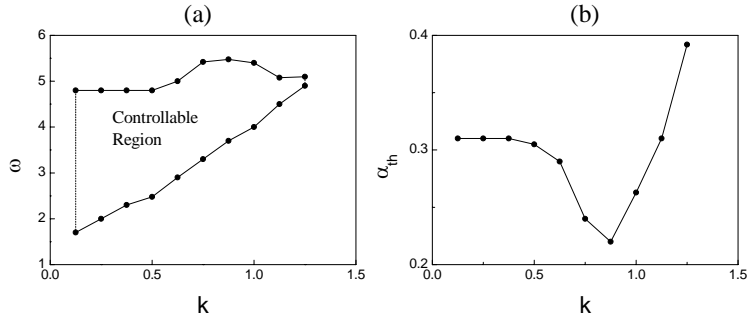

Fig.3. (a) Controllable region in the plane of the control parameters k and $\omega$ for $\alpha = 0.32$ and (b) threshold of the control amplitude versus k for $\omega = 4.6$ in the 1D case. $\varepsilon = 0.1$, $M = 10$.

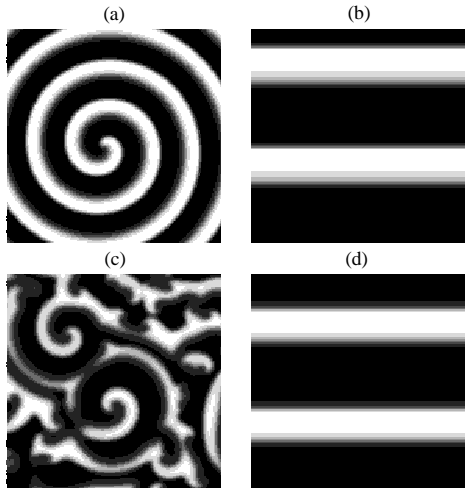

Fig.4. Gray scale plot of E in a region of $27 \times 27$ in the 2D case. White and dark regions correspond to excited and recovery regions, respectively. The left column (without control) are (a) the spiral wave for $\varepsilon = 0.05$ and $M = 4$ and (c) the turbulent state for $\varepsilon = 0.1$ and $M = 10$. The right column (with control) are (b) the travelling wave from (a), with $k = 0.5$, $\omega = 5$, and $\alpha = 0.26$, and (d) the travelling wave from (c) with $k = 0.5$, $\omega = 4$, and $\alpha = 0.29$.

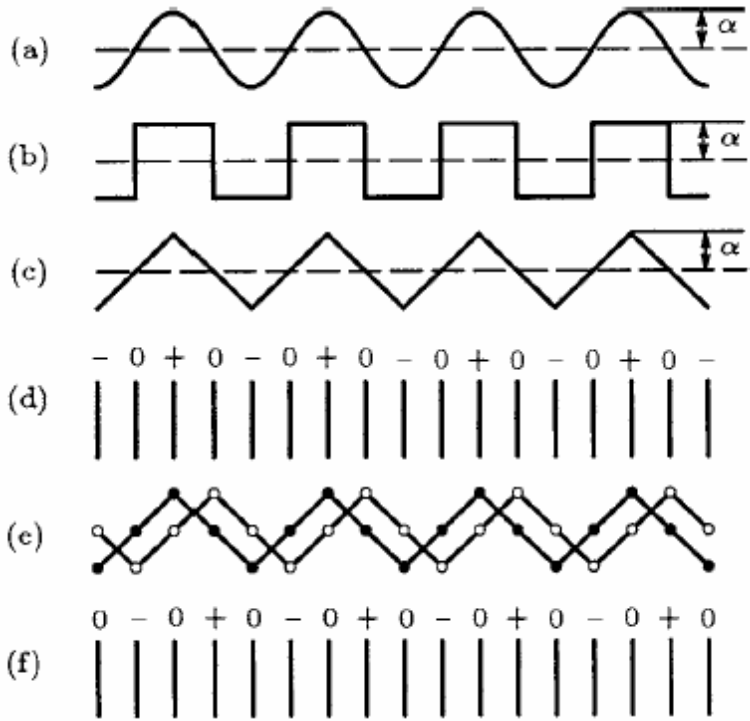

Fig.5. Control signals of (a) sinusoidal, (b) square, and (c) triangular travelling waves where the dashed lines indicate the zero-voltage levels. All the control signals are symmetric with respect to the zero-voltage lines. α indicates the amplitude of the waves. (d) Voltage on each electrode at one moment t. (e) Spatial voltage distribution curves (solid dots and circles corresponding to time t and t +Δt, respectively). (f) Voltage on each electrode at next time step t +Δt.